# Multi-domain Polarization Switching in Hf$_{0.5}$Zr$_{0.5}$O$_2$-Dielectric Stack: The Role of Dielectric Thickness

Atanu K. Saha[1,a)], Mengwei Si[1], Peide D. Ye[1], and Sumeet K. Gupta[1]

[1]School of Electrical and Computer Engineering, Purdue University, West Lafayette, IN 47907, US

a)Author to whom correspondence should be addressed: saha26@purdue.edu

**ABSTRACT**

**We investigate the polarization switching mechanism in ferroelectric-dielectric (FE-DE) stacks and its dependence on the dielectric thickness ($T_{DE}$). We fabricate HZO-Al$_2$O$_3$ (FE-DE) stack and experimentally demonstrate a decrease in remnant polarization and an increase in coercive voltage of the FE-DE stack with an increase in $T_{DE}$. Using phase-field simulations, we show that an increase in $T_{DE}$ results in a larger number of reverse domains in the FE layer to suppress the depolarization field, which leads to a decrease in remanent polarization and an increase in coercive voltage. Further, the applied voltage-driven polarization switching suggests domain-nucleation dominant characteristics for low $T_{DE}$, and domain-wall motion-induced behavior for higher $T_{DE}$. In addition, we show that the hysteretic charge-voltage characteristics of the FE layer in the FE-DE stack exhibit a negative slope region due to the multi-domain polarization switching in the FE layer. Based on our analysis, the trends in charge-voltage characteristics of the FE-DE stack with respect to different $T_{DE}$ (which are out of the scope of single-domain models) can be described well with multi-domain polarization switching mechanisms.**

Ferroelectric (FE) hafnium oxide, by virtue of its CMOS process compatibility[1,2] and rich domain dynamics[3,4], has been identified as one of the most promising candidates for enabling future electronic devices. By integrating doped HfO$_2$ (as FE) in the gate stack of a transistor (FE-FET), non-volatile memory (NVM)[5-6], neuron[7] and synaptic[8,9] functionalities have been demonstrated. Such diverse functionalities demand different characteristics of polarization ($P$) switching in the FE layer. For example, an abrupt $P$-switching is beneficial for neurons and binary NVMs, while a gradual $P$-switching is favorable for multi-bit memories and synapses. Therefore, it becomes essential to appropriately design FEFET for application-specific device behavior, for which gate stack optimization plays a key role.

In the FEFET gate stack, a dielectric (DE) layer exists between the FE and the semiconductor channel[5-10], which can significantly impact the FEFET characteristics[11]. According to the single-domain Landau-Khalatnikov (LK) model of FE, an increase in DE thickness ($T_{DE}$) should increase the depolarization field and reduce the coercive voltage ($V_C$) of the FE-DE stack[12,13]. However, the FE-DE stack with Zr-doped HfO$_2$ (HZO) as the FE and Al$_2$O$_3$/HfO$_2$ as the DE have been demonstrated[14-15] to exhibit an increasing coercive voltage ($V_C$) with the increase in $T_{DE}$. Therefore, it is important to bridge the gap between the theoretical understanding and experimental observations regarding the role of $T_{DE}$ in FEFET and for that, a common approach is to analyze the FE-DE stack[12-13]. To that end, in this letter, we experimentally and theoretically analyze the $P$-switching in FE-DE stack with HZO as the FE and Al$_2$O$_3$ as the DE layer. Our results signify an increase in $V_C$, a decrease in remanent-$P$ ($P_R$) and a decrease in $P$-switching slope with the increase in $T_{DE}$. By employing phase-field simulations, we show that such dependencies can be attributed to the multi-domain phenomena in FE[16] which cannot be captured in the single-domain $P$-switching model. To further study the role of the dielectric, we analyze the dependence of $V_C$, $P_R$ and the switching slope of the FE-DE stack on the dielectric permittivity through phase-field simulations.

For the fabrication of the FE-DE stacks, we start with the standard solvent cleaning of heavily p-doped Si substrates. Then, 30nm TiN layer is deposited by atomic layer deposition (ALD) at 250 °C, using [(CH$_3$)$_2$N]$_4$Ti and NH$_3$ as the Ti and N precursors, respectively. After this, an HZO film is deposited by ALD at 200 °C, using [(CH$_3$)$_2$N]$_4$Hf, [(CH$_3$)$_2$N]$_4$Zr, and H$_2$O as the Hf, Zr, and O precursors, respectively. HfO$_2$:ZrO$_2$ cycle ratio of 1:1 is used to form the 10nm Hf$_{0.5}$Zr$_{0.5}$O$_2$ film. Similarly, on top of HZO, an Al$_2$O$_3$ layer is deposited followed by a 30nm TiN layer deposition. After that, the samples are annealed at 500 °C in N$_2$ environment for 1 minute by rapid thermal annealing. Then, Ti/Au top electrodes are fabricated using photolithography, e-beam evaporation, and lift-off process (area=5024μm$^2$). The average polarization ($P_{avg}$) versus applied voltage ($V_{app}$) measurement is carried out using a Radiant RT66C FE tester at room temperature at a very low frequency (50Hz). Considering the polarization switching time in HZO (<1μs)[17], such low-frequency measurements can be considered as quasi-static. Fig. 1(a) shows the $P_{avg}$-$V_{app}$ characteristics of FE-DE stack for 10nm HZO and 1/3/5nm Al$_2$O$_3$. Our results show a decrease in $P_R$ ($P_{avg}$ at $V_{app}$=0V), an increase in $V_C$ ($V_{app}$ at $P_{avg}$=0) and a decrease in $P$-switching slope with the increases in $T_{DE}$. To explain such dependencies, we now analyze the $P$-switching in the FE-DE stack based on multi-domain phase-field simulation.

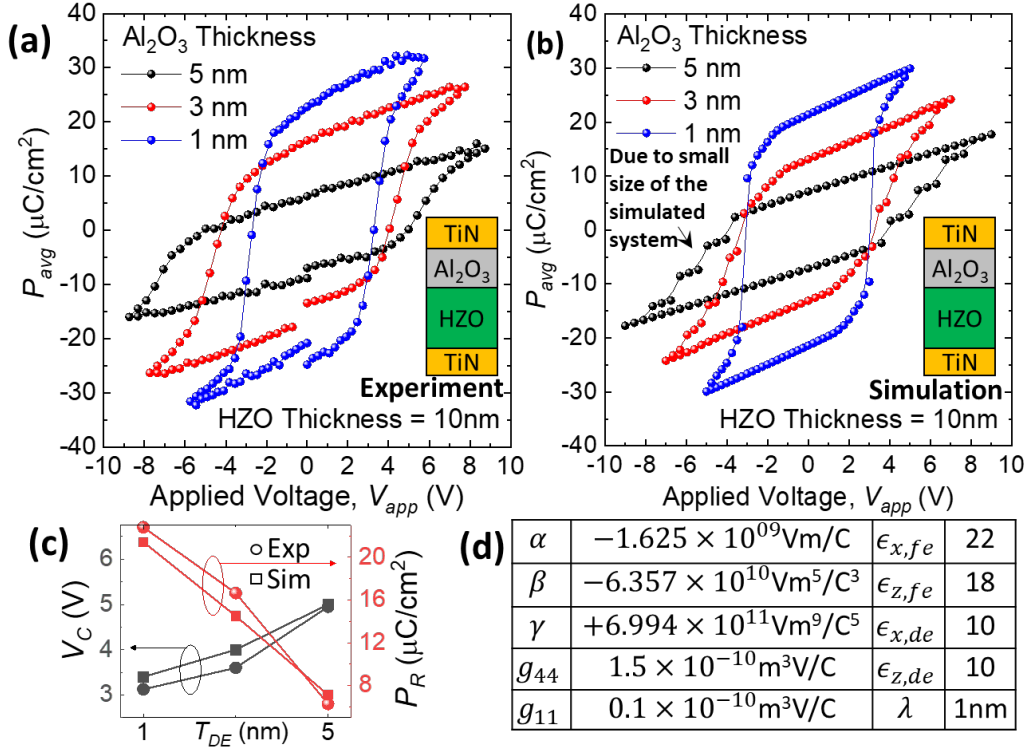

Figure 1: (a) Measured and (b) simulated $P_{avg}$-$V_{app}$ characteristics of FE-DE stack for different $T_{DE}$. (c) average $V_C$ ($V_{app}$ at $P_{avg}$=0) and $P_R$ ($P_{avg}$ at $V_{app}$=0) for different $T_{DE}$. (d) Table showing the simulation parameters.

In our 2D phase-field simulation, we self-consistently solve the time-dependent Ginzburg-Landau (TDGL) equation (eqn. 1) and Poisson's equation (eqn. 2) as shown below[18,19].

$$-\frac{1}{\Gamma}\frac{\partial P}{\partial t} = \alpha P + \beta P^3 + \gamma P^5 - g_{11}\frac{d^2P}{dz^2} - g_{44}\frac{d^2P}{dx^2} + \frac{d\phi}{dz} \quad (1)$$

$$-\epsilon_0\left[\frac{\partial}{\partial x}\left(\epsilon_x \frac{\partial \phi}{\partial x}\right) + \frac{\partial}{\partial z}\left(\epsilon_z \frac{\partial \phi}{\partial z}\right)\right] = -\frac{dP}{dz} \quad (2)$$

Here, $\alpha$, $\beta$, and $\gamma$ are Landau coefficients; $g_{11(44)}$ is the gradient coefficient; $\epsilon_{z(x)}$ is relative background permittivity; $\Gamma$ is viscosity coefficient; $\phi$ is potential; $P$ is the polarization of FE unit cell. We assume that the $P$-direction (c-axis of the orthorhombic HZO crystal) is parallel to the film thickness (z-axis)[18,19]. Note that the $dP/dz$ induces charges in the FE layer and thus enters in eqn. 2. At the FE-DE interface, $\lambda(dP/dz)$-$P$=0 is used for the surface energy contribution, where $\lambda$ is the extrapolation length[20,21]. All simulation parameters are given in Fig. 1(d). Due to the noncentrosymmetric crystal and lower elastic interactions in the out-of-plane direction compared to the in-plane direction in HZO[22], we use $g_{11}$<$g_{44}$. Similarly, as the $P$-direction is along the $z$-axis, therefore, a lower number of atoms per unit cell take part in $\epsilon_z$ compared to $\epsilon_x$ and hence, $\epsilon_z$<$\epsilon_x$ (which is similar to other FE like PZT[23]). We consider the length ($l$, along the x-direction) of the system to be 30nm which is analogous to the average grain size of HZO[24]. To be consistent with the experimental measurements, the simulations are performed based on the quasi-static criteria (negligible $dP/dt$). Therefore, our simulation results are independent of the value of $\Gamma$. Further, we use a smaller FE region (equivalent to the size of a grain ~30nm) in simulation compared to the area of our experimental sample because of the scale-free nature of the FE HZO[22]. Thus, our simulations capture the trends with respect to the mean behavior of a single grain; however, for capturing the effects such as variation in coercive fields, a multi-grain simulation is needed which is out of the scope of this work. In the multi-domain scenario, $P_{avg}$ is computed by integrating the displacement field at the metal-DE (or metal-FE) interface ($P_{avg}$= ($\int \epsilon_0 \epsilon_{z,DE} E_{z,DE} dx)/l$ = ($\int (P + \epsilon_0 \epsilon_{z,FE} E_{z,FE}) dx)/l$). Here, the $E_{z,FE(DE)}$ is the out-of-plane (z) component of the electric field in the FE (DE) layer. The simulated $P_{avg}$-$V_{app}$ characteristics of the FE-DE stack is shown in Fig. 1(b) illustrating a good agreement with the experiments (Fig. 1(c)). The mismatch in the $P$-switching region can be reduced by simulating multiple grains (discussed later).

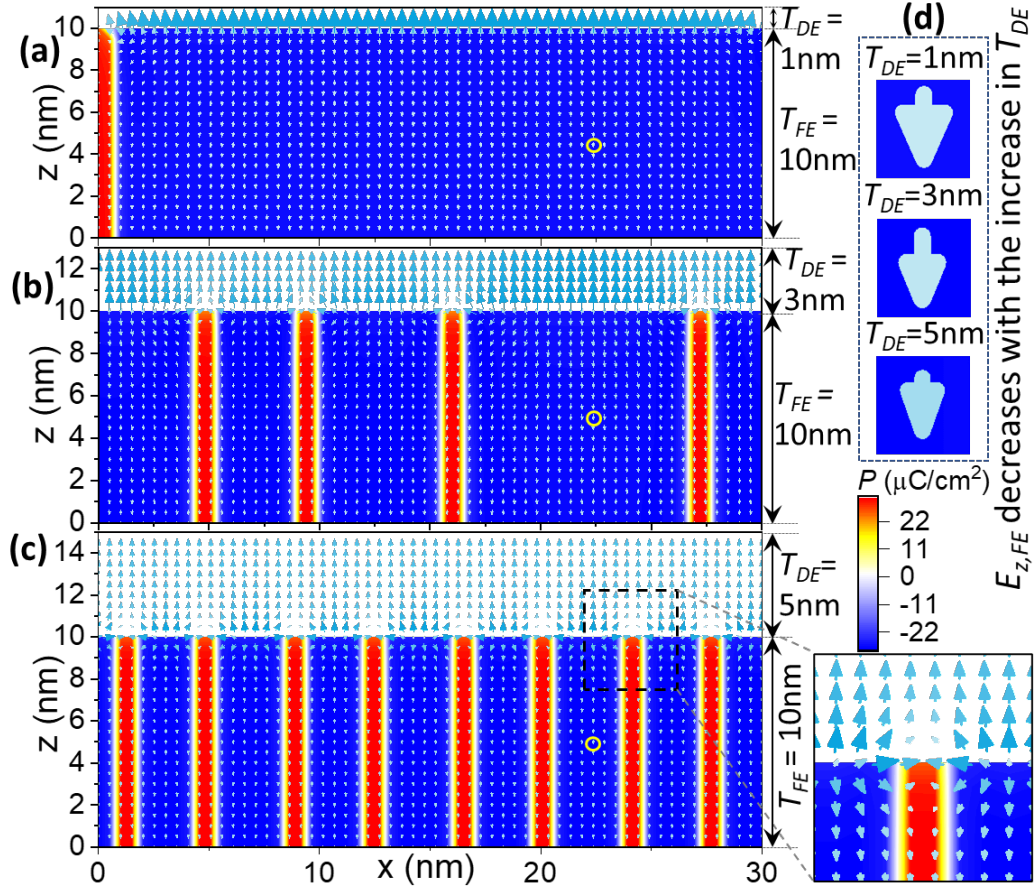

Figure 2: Simulated $P$ (color map) and $E$-field (arrow) profile in FE-DE for $T_{DE}$ = (a) 1nm (b) 3nm, and (c) 5nm. The blue (red) regions signifying -$P$ (+$P$) domains. (d) $E_{z,FE}$ in FE at the yellow circle shown in (a-c).

To explain these characteristics, let us start with $V_{app}$=0V and $P_R$ <0. In an FE-DE stack, the $P$-induced bound charges appear near the FE-DE interface leading to a non-zero $E_{z,DE}$ and $E_{z,FE}$. If $P$ is homogeneous (e.g. in single-domain (SD) state), then $E_{z,FE}$ will be directed opposite to the $P$-direction yielding depolarization energy $f_{dep}$ (= $-PE_{z,FE}$). At the same time, $E_{z,FE}$ will reduce the $P$ magnitude (|$P$|) leading to an increase in the free energy ($f_{free}$). In order to suppress $f_{dep}$ and $f_{free}$ (to minimize the overall energy), FE breaks into multiple domains with opposite $P$-directions. In this multi-domain (MD) state, the $P$-induced bound charges at the FE-DE interface not only give rise to $E_{z,FE(DE)}$ (as before), but also form in-plane $E$-field ($E_{x,FE(DE)}$) called stray field[18,25]. As a portion of the bound charge gets compensated by the stray-field, $E_{z,FE(DE)}$ is reduced in the MD state (compared to the SD state) leading to a reduction in $f_{dep}$ and $f_{free}$. However, this suppression of $f_{dep}$ and $f_{free}$ occurs at the cost of (i) gradient energy, $f_{grad}$ (=$g_{44}(dP/dx)^2$) due to the spatial variation of $P$ in the domain-walls (DW) and (ii) electrostatic energy $f_{elec}$ (=$\epsilon_0\epsilon_{x,FE}E_{x,FE}^2$) due to the stray fields. Hence, the formation of the MD state occurs as an interplay among competing energy components to obtain the minimum energy. With this understanding, let us now discuss the impact of $T_{DE}$ on $P_R$.

In the FE-DE stack (at $V_{app}$=0 and $P_R$<0), an increase in $T_{DE}$ tends to increase $E_{z,FE}$ due to the higher voltage drop across DE and an equal and opposite voltage drop across the FE layer. This increase in $E_{z,FE}$ tends to increase $f_{dep}$ and $f_{free}$. To counter this, a larger number of oppositely polarized domains (+$P$ in Fig. 2) appear that create more stray fields to suppress $E_{z,FE}$. The simulated $P$ and $E$-field profiles in Fig. 2(a-c) validate the increase in the number of +$P$ domains (red domains) and suppression of $E_{z,FE}$ (Fig. 2(d)) with the increase in $T_{DE}$. The appearance of a larger number of +$P$ domains leads to a smaller size of -$P$ domains (blue) and hence, reduced |$P_R$| with the increase in $T_{DE}$ (Fig. 1(a-c)).

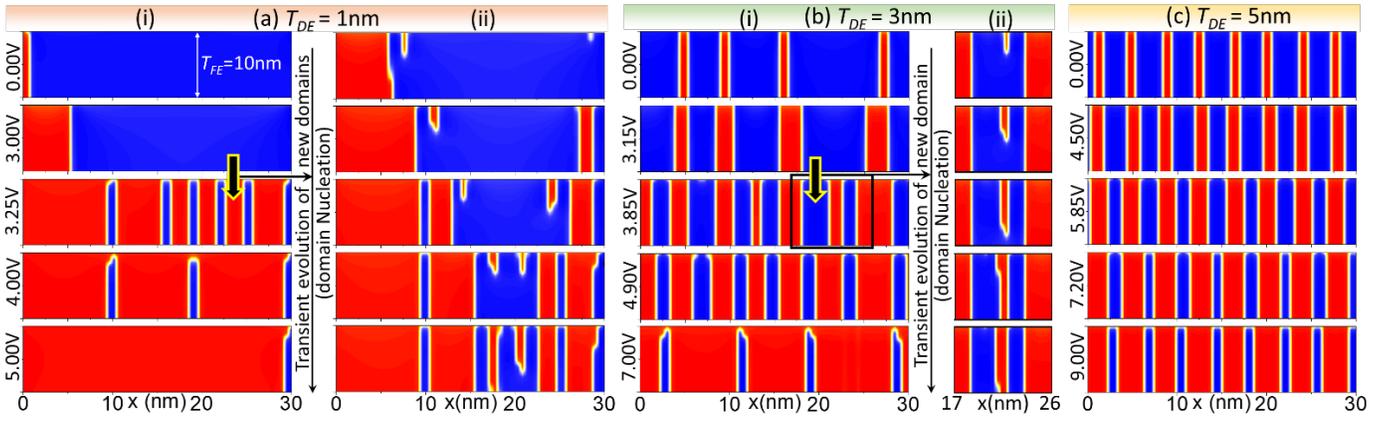

Figure 3: Simulated polarization profile in FE at a different applied voltage ($V_{app}$) in FE-DE stack for different $T_{DE}$ = (a) 1nm, (b) 3nm and (c) 5nm showing domain nucleation and domain-wall motion based polarization switching. In all the cases, the FE thickness is 10nm.

Now, let us discuss $V_{app}$-induced $P$-switching. $P$-switching can take place if $f_{grad}+f_{dep}+f_{elec}+f_{free} > \max(f_{free})$. In the MD state, $E_{z,FE}$ is maximum away from DW near the FE-DE interface, which leads to maximum $f_{dep}$. In contrast, $f_{grad}$ is maximum near the DW due to the largest variation in $P$. Now, with an increase in $V_{app}$, $E_{z,FE}$ increases leading to a change in $P$ magnitude (|$P$| increases in +$P$ domains and decreases in -$P$ domains). Thus, $f=f_{grad}+f_{dep}+f_{elec}+f_{free}$ increases[16]. If the increase in $f$ is dominant near the DW, then $P$-switching occurs through DW motion. However, if the increase in $f$ is dominant away from the DW, then $P$-switching occurs through the nucleation of new domains. $P$ profiles at different $V_{app}$ are shown in Fig. 3(a)-i for $T_{DE}$=1nm. With the increase in $V_{app}$, $P$-switching starts through DW motion (at $V_{app}$=1.5V) and at $V_{app}$>3V several new domains nucleate causing a denser domain pattern. The transient nature of domain nucleation is shown in Fig. 3(a)-ii signifying their formation starting from the FE-DE interface. Once, the domain pattern becomes denser, a significant portion of $E_{z,FE}$ is suppressed by the stray fields at the expense of an increased $f_{grad}$. Hence, with further increase in $V_{app}$, $P$-switching takes place through DW motion leading to complete switching of several domains. Similarly, for $T_{DE}$=3nm (Fig. 3(b)-i-ii), $P$-switching initiates through DW motion (at $V_{app}$=1.85V) followed by domain nucleation (at $V_{app}$>3.15V) and then DW motion. However, for $T_{DE}$=5nm (Fig. 3(c)), the initial domain pattern is much denser, which suppresses $E_{z,FE}$ at the cost of $f_{grad}$. Hence, nucleation of new domains is not observed, and $P$-switching takes place only through DW motion (at $V_{app}$>3.6V).

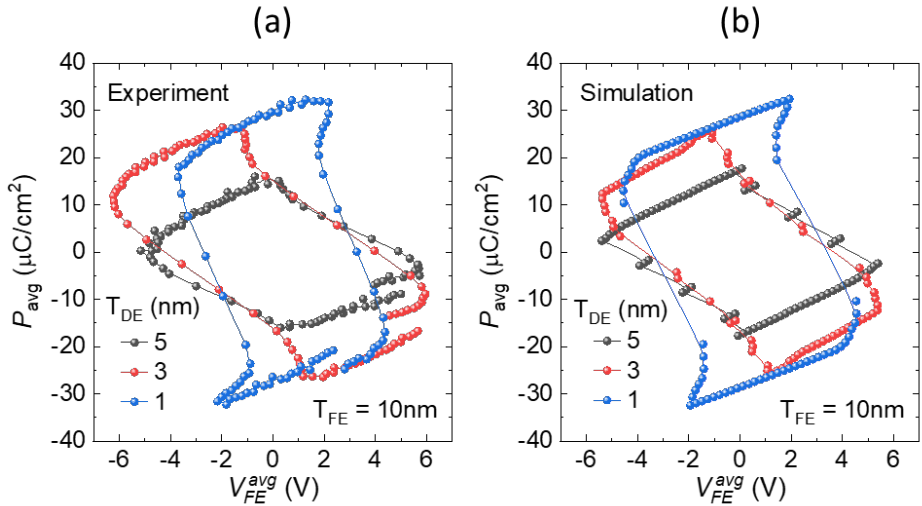

Figure 4: Extracted $P_{avg}$-$V_{FE}$ characteristics of FE in the FE-DE stack from the (a) experimental and (b) simulated $P_{avg}$-$V_{app}$ characteristics.

The extracted $P_{avg}$-$V_{FE}$ characteristics from experimental and simulated $P_{avg}$-$V_{app}$ characteristics are shown in Fig. 4 (a-b) signifying the negative $dP_{avg}/dV_{FE}$ region. Such negative $dP_{avg}/dV_{FE}$ exists during the $P$-switching in the FE layer via domain nucleation and/or DW motion. Recall that the local $E_{z,FE}$ in the FE layer is depolarizing i.e. opposite to the direction of $P$. Now, let us consider the FE-DE stack is in the $P_R<0$ state (average $E_{z,FE}>0$). When $V_{app}$ is increased and leads to MD P-switching, the $P_{avg}$ increases either through the formation of new +P domains (nucleation) or through the size increase of +P domains (DW displacement). Both of these phenomena lead to a decrease in average $E_{z,FE}$ (i.e. the average $E_{z,FE}$ becomes less positive). As the increase in $P_{avg}$ accompanies the decrease in average $V_{FE}$ (=$T_{FE}E_{z,FE}$), thus the $dP_{avg}/dV_{FE}$ becomes negative. A similar effective negative capacitance effect occurs for $P_R>0$. It is important to note that the appearance of negative $dP_{avg}/dV_{FE}$ is an electrostatic effect rather than a transient artifact. However, the actual slope of $dP_{avg}/dV_{FE}$ can certainly be impacted by the frequency of the applied $V_{app}$ due to the time-dependency of domain-nucleation and DW-motion.

The DW motion occurs via lattice-by-lattice propagation yielding a gradual increase in $P_{avg}$. However, the nucleation of a new domain involves simultaneous P-switching in several lattices leading to a sharper change in $P_{avg}$. Since, with an increase in $T_{DE}$, the dominant P-switching mechanism changes from nucleation to DW-motion-based, P-switching becomes more gradual - Fig. 1(a-b)). Further, our simulations show a step-wise P-switching behavior for $T_{DE}$=5nm (Fig. 1(b)), where each step jump signifies the DW displacement, and the flatter region corresponds to no DW displacement. The non-zero slope of the flat region is due to the response of P magnitudes and $\epsilon_{z,FE}$ to $E_{z,FE}$. In this flat region, with an increase in $V_{app}$, $E_{z,FE}$ first increases. If the increase in $E_{z,FE}$ is beyond a critical value so that $f>max(f_{free})$, then the P-switching takes place via DW displacement. Recall that the P-switching leads to an increase in $P_{avg}$ and a simultaneous reduction in $E_{z,FE}$. This yields a negative slope in the $P_{avg}$-$V_{FE}$ characteristics ($dP_{avg}/dV_{FE}<0$) and a step jump in the $P_{avg}$-$V_{app}$ characteristics. Now, after each P-switching step, to induce further DW motion, $V_{app}$ needs to be increased to increase $E_{z,FE}$ beyond a (new) critical value. Consequently, we observe a step-wise P-switching behavior in Fig. 1(b) and Fig 4(b). However, such step-jumps are absent in the measured characteristics because of the larger area (lots of grains) of the fabricated sample compared to our simulation (~one grain). Thus, even though the DW motion may absent in some of the grains of the experimental sample, it may be present in other grains (due to the variation in grain size and/or crystallographic angle) leading to a continuous increase in $P_{avg}$. Hence, we expect that simulation of a larger system considering multiple grains may reduce this mismatch between the simulation and experimental results.

Let us now explain the effect of $T_{DE}$ on $V_C$ (defined as the $V_{app}$ where $P_{avg}=0$). For that, we consider the voltage drop across FE averaged along the length (referred as $V_{FE}=(\iint E_{z,FE}dxdz)/l$). As an increase in $T_{DE}$ suppresses $E_{z,FE}$ (discussed before), it leads to a decrease in $V_{FE}$ at $V_{app}$=0V. With the decrease in initial $V_{FE}$, a higher $V_{app}$ is required to achieve a critical $V_{FE}$ to trigger P-switching. Therefore, the DW motion initiates at $V_{app}$=1.5V for $T_{DE}$=1nm and at 1.85V for $T_{DE}$=3nm. Similarly, the domain nucleation takes place at $V_{app}$>3V for $T_{DE}$=1nm and $V_{app}$>3.15V for $T_{DE}$=3nm. Further, $dP_{avg}/dV_{app}$ decreases with the decrease in $T_{DE}$ (discussed before) leading to an increase in required $V_{app}$ to achieve $P_{avg}$=0. Due to the decrease in initial $V_{FE}$ (at $V_{app}$=0V) and lower $dP_{avg}/dV_{app}$, $V_C$ of the FE-DE stack increases with an increase in $T_{DE}$. Note that the increase in $V_C$ for larger $T_{DE}$ cannot be captured by the SD mode, but can be described well considering the MD effects (as explained above).

So far we have discussed different attributes of $P_{avg}$-$V_{app}$ characteristics of FE-DE stack with respect to different $T_{DE}$. This can also be regarded as the equivalent of different DE capacitances, $C_{DE}=\epsilon_0\epsilon_{DE}/T_{DE}$. Now, one may argue that the $P_{avg}$-$V_{app}$ characteristics can be also tuned by using a different DE material ($\epsilon_{DE}$). While this is indeed possible (and can be an important design knob), the effect of $\epsilon_{DE}$ is not just changing $C_{DE}$, but involves some more physical processes that mandate further analysis. To decouple the effect of $\epsilon_{DE}$ on $C_{DE}$, we theoretically analyze the dependence of FE-DE characteristics on $\epsilon_{DE}$ in the supplementary sections by simultaneously and proportionally changing $T_{DE}$ to keep the same $C_{DE}$. We show that the $P_{avg}$-$V_{app}$ characteristics are not unique to $C_{DE}$, rather depends on the choice of $\epsilon_{DE}$. Such dependency originates due to the electrostatic boundary condition of the in-plane electric field at the FE-DE interface. Considering different $\epsilon_{DE}$ (but same $C_{DE}$), our simulation results suggest that the $V_C$ decreases and $P_R$ increases with the decrease in $\epsilon_{DE}$. We discuss such $\epsilon_{DE}$ dependency on the P-switching in FE-DE stack in the supplementary section.

In summary, we show that the FE layer forms a denser domain pattern with increasing $T_{DE}$ by suppressing the depolarization field and leading to a higher hysteresis in the FE-DE stack. Simultaneously, the mechanism of P-switching can be modulated from nucleation to DW-motion dominant by increasing $T_{DE}$. In addition, we show that the DW energy and thus the coercive voltage and remanent P can further be modulated by $\epsilon_{DE}$ while keeping the same $C_{DE}$. Such $T_{DE}$ and $\epsilon_{DE}$ dependency can serve as the potential knobs to deploy the application-driven optimization of the FEFET gate stack. For instance, FEFETs with low $T_{DE}$ (high switching slope) can be used for the design of binary NVMs and neurons, while high $T_{DE}$ can be utilized for multi-bit memories and synapse designs.

## ACKNOWLEDGEMENT

This work was supported in part by Semiconductor Research Corporation (SRC) under contract no. 2020-LM-2959 and National Science Foundation (NSF) under grant no. 1814756 and grant no. 2008412.

## DATA AVAILABILITY

The data that support the findings of this study are available from the corresponding author upon reasonable request.

## REFERENCES

[1] J. Müller, T. S. Böscke, S. Müller, E. Yurchuk, P. Polakowski, J. Paul, D. Martin, T. Schenk, K. Khullar, A. Kersch, W. Weinreich, S. Riedel, K. Seidel, A. Kumar, T. M. Arruda, S. V. Kalinin, T. Schlösser, R. Boschke, R. van Bentum, U. Schröder, and T. Mikolajick, *IEDM Tech. Dig.*, pp. 10.8.1–10.8.4, Dec. 2013. DOI: 10.1109/IEDM.2013.6724605
[2] A. K. Saha, B. Grisafe, S. Datta, and S. K. Gupta, *Proc. IEEE VLSI Technol.*, pp. T226-T227, Jun. 2017. DOI: 10.23919/VLSIT.2019.8776533
[3] K. Ni, B. Grisafe, W. Chakraborty, A. K. Saha, S. Dutta, M. Jerry, J. A. Smith, S. Gupta, and S. Datta, *IEDM Tech. Dig.*, pp. 16.1.1-16.1.4, Dec. 2018. DOI: 10.1109/IEDM.2018.8614527
[4] A. K. Saha, K. Ni, S. Dutta, S. Datta, and S. Gupta, *Appl. Phys. Lett.*, vol. 114, no. 20, pp. 20903, May. 2019. DOI: 10.1063/1.5092707
[5] K. Chatterjee, S. Kim, G. Karbasian, A. J. Tan, A. K. Yadav, A. I. Khan, C. Hu, and S. Salahuddin, *IEEE Electron Device Lett.*, vol. 38, no. 10, pp. 1379-1382, Oct. 2017. DOI: 10.1109/LED.2017.2748992
[6] S. Dünkel, M. Trentzsch, R. Richter, P. Moll, C. Fuchs, O. Gehring, M. Majer, S. Wittek, B. Müller, T. Melde, H. Mulaosmanovic, Stefan Slesazeck, S. Müller, J. Ocker, M. Noack, D-A Löhr, P. Polakowski, J. Müller, T. Mikolajick, J. Höntschel, B. Rice, J. Pellerin, S. Beyer, *IEDM Tech. Dig.*, pp. 19.7.1-19.7.4, Dec. 2017. DOI: 10.1109/IEDM.2017.8268425
[7] H. Mulaosmanovic, E. Chicca, M. Bertele, T. Mikolajickac, and S. Slesazecka, *Nanoscale*, vol. 10, no. 46, pp. 21755-21763, Nov. 2018. DOI: 10.1039/C8NR07135G
[8] H. Mulaosmanovic, J. Ocker, S. Müller, M. Noack, J. Müller, P. Polakowski, T. Mikolajick, and S. Slesazeck, *IEEE VLSI Technol.*, pp. T176-T177, Jun. 2017. DOI: 10.23919/VLSIT.2017.7998165
[9] M. Jerry, P. Chen, J. Zhang, P. Sharma, K. Ni, S. Yu, and S. Datta, *IEDM Tech. Dig.*, pp. 6.2.1-6.2.4, Dec. 2017. DOI: 10.1109/IEDM.2017.8268338
[10] P. Sharma, K. Tapily, A. K. Saha, J. Zhang, A. Shaughnessy, A. Aziz, G. L. Snider, S. Gupta, R. D. Clark, and S. Datta, *IEEE VLSI Technol.*, pp. T154–T155, Jun. 2017. DOI: 10.23919/VLSIT.2017.7998160
[11] V. Gaddam, D. Das, and S. Jeon, *IEEE Trans. on Electron Devices*, vol. 67, no. 2, pp. 745-750, Feb. 2020. DOI: 10.1109/TED.2019.2961208)
[12] S. Salahuddin, and S. Datta, *Nano Lett.*, vol. 8, no. 2, pp. 405-410, Dec. 2007. DOI: 10.1021/nl071804g
[13] A. K. Saha, S. Datta, and S. Gupta, *J. Appl. Phys.*, vol. 123, no. 10, pp. 105102, Mar. 2018. DOI: 10.1063/1.5016152
[14] W. Xiao , C. Liu, Y. Peng , S. Zheng , Q. Feng , C. Zhang ,J. Zhang , Y. Hao, M. Liao, and Y. Zhou, *IEEE Electron Device Lett.*, vol. 40, no. 5, pp. 714-717, Mar. 2019. DOI: 10.1109/LED.2019.2903641
[15] M. Si, X. Lyu, and P. D. Ye, *ACS Appl. Electron. Mater.*, vol. 1, no. 5, pp. 745-751, May. 2019. DOI: 10.1021/acsaelm.9b00092
[16] A. M. Bratkovsky and A. P. Levanyuk, "Abrupt appearance of the domain pattern and fatigue of thin ferroelectric films," *AIP Conference Proceedings*, vol. 535, no. 1, pp. 218-228, Sep. 2000. DOI: 10.1063/1.1324458
[17] M. Si, X. Lyu, P. R. Shrestha, X. Sun, H. Wang, K. P. Cheung, and P. D. Ye, *Appl. Phys. Lett.*, vol. 115, no. 7, pp. 072107, Aug. 2019. DOI: 10.1063/1.5098786
[18] A. K. Saha, and S. K. Gupta, *Scientific Reports*, vol. 10, no. 1, pp. 10207, Jun. 2020. DOI: 10.1038/s41598-020-66313-1
[19] H.W. Park, J. Roh, Y. B. Lee, and C. S. Hwang, *Adv. Mater.*, vol. 31, no. 32, pp. 1805266, Jun. 2019. DOI: 10.1002/adma.201805266
[20] M. D. Glinchuk and E. A. Eliseev, *J. Appl. Phys.*, vol. 93, no. 2, pp. 1150, Dec. 2003. DOI: 10.1063/1.1529091
[21] P. Chandra and P. B. Littlewood, *Physics of Ferroelectrics: A Modern Perspective, Topics Appl. Physics*, vol. 105, K. Rabe, C. H. Ahn, J.-M. Triscone, Eds., Berlin: Springer-Verlag, 2007, pp. 69-116. DOI: 10.1007/978-3-540-34591-6
[22] H.-J. Lee, M. Lee, K. Lee, J. Jo, H. Yang, Y. Kim, S. C. Chae, U. Waghmare, J. H. Lee, *Science*, vol. 369, no. 6509, pp. 1343-1347, Sep. 2020. DOI: 10.1126/science.aba0067
[23] A. I. Kurchak, E. A. Eliseev, S. V. Kalinin, M. V. Strikha, and A. N. Morozovska, *Phys. Rev. Applied*, vol. 8, no. 2, pp. 024027, Aug. 2017.DOI: 10.1103/PhysRevApplied.8.024027
[24] H. J. Kim, M. H. Park, Y. J. Kim, Y. H. Lee, W. Jeon, T. Gwon, T. Moon, K. D. Kim, and C. S. Hwanga, *Appl. Phys. Lett.*, vol. 105, no. 19, pp. 192903, Nov. 2014. DOI: 10.1063/1.4902072
[25] J. Íñiguez, P. Zubko, I. Luk'yanchuk, and A. Cano, *Nat. Rev. Mater.*, vol. 4, no. 4, pp. 243–256, Mar. 2019. DOI: 10.1038/s41578-019-0089-0

# Supplementary Section

# Multi-domain Polarization Switching in Hf$_{0.5}$Zr$_{0.5}$O$_2$-Dielectric Stack: The Role of Dielectric Thickness

Atanu K. Saha, Mengwei Si, Peide D. Ye, and Sumeet K. Gupta

School of Electrical and Computer Engineering, Purdue University, West Lafayette, IN 47907, US

*Effects of $\epsilon_{DE}$ in the P-switching characteristics in the FE-DE stack:*

In this section, we discuss the effect of different DE permittivity ($\epsilon_{DE}$) on the P switching characteristics of FE-DE stack while we keep the capacitance of the DE layer same (iso-$C_{DE}=\epsilon_0\epsilon_{DE}/T_{DE}$). First, it is important to note that the electrostatic boundary conditions at the FE-DE interface need to satisfy not only the continuity of the out-of-plane displacement field ($D_{z,FE} = D_{z,DE}$) but also the equality of the in-plane electric-field ($E_{x,FE} = E_{x,DE}$). The in-plane interface E-field ($E_x$) between two opposite polarization domains (+P and -P) follows $|P|=\epsilon_0\epsilon_x|E_x|$ and that implies an increase in $E_x$ with the decrease in in-plane DE permittivity ($\epsilon_{x,DE}$). That, in turn, leads to an enhanced $E_{x,FE}$ and causes an increase in the in-plane electrostatic energy ($\epsilon_{x,FE}E_{x,FE}^2$) in the FE layer. Our simulation suggests that such an increase in in-plane electrostatic energy is partially compensated by reducing the $P$ magnitude. However, this happens at the cost of an increase in the $f_{free}$. Overall, this leads to an increase in the local energy ($f$) near the DW for lower $\epsilon_{DE}$ (at iso-$C_{DE}$). This would imply that a smaller applied field (or $V_{app}$) will be required to initiate the P-switching in the FE-DE stack. Considering different $\epsilon_{DE}$, our simulated $P_{avg}$-$V_{app}$ characteristics are shown in Fig. S1 signifying that the DW motion initiates at a lower voltage with the decrease in $\epsilon_{DE}$. Further, with the increase in coercive voltage of DW motion, the polarization switching via DW motion becomes more prominent for the lower $\epsilon_{DE}$ and same $V_{app}$. Consequently, the $P_R$ increases with the decrease in $\epsilon_{DE}$. It is important to note that the effect of $\epsilon_{DE}$ is more prominent in the case of lower $C_{DE}$ (Fig. S1). This is because for lower $C_{DE}$ (or higher $T_{DE}$) the P-switching mechanism is dominated by DW displacement and thus the change in the DW energy is more significant for lower $C_{DE}$. Note that such $\epsilon_{DE}$ dependency can be used as a potential knob to optimize the write voltage and retention of the FEFETs for NVM application.

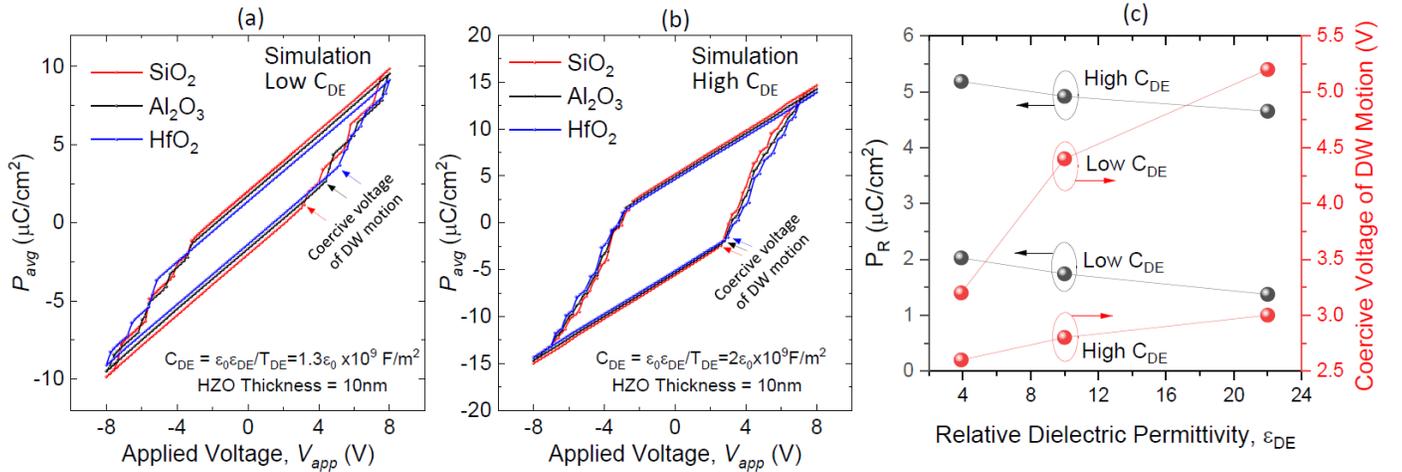

Figure S1: $P_{avg}$-$V_{app}$ characteristics of FE-DE stack with different relative DE permittivity, $\epsilon_{DE}$ and iso-$C_{DE}$ (=$\epsilon_0\epsilon_{DE}/T_{DE}$) considering (a) $C_{DE} = 1.3\epsilon_0 \times 10^9\ F/m^2$ (Low) and (b) $C_{DE} = 2\epsilon_0 \times 10^9\ F/m^2$ (High). The coercive voltage of DW motion is shown as the arrow in (a-b). (c) The coercive voltage of DW motion and remanent $P_{avg}$ ($P_R$) with respect to $\epsilon_{DE}$ for the low and high $C_{DE}$ in Fig. S1 (a-b).